\documentclass[journal]{IEEEtran}

\usepackage{graphicx}
\usepackage{amsmath}
\usepackage{mathtools}
\usepackage[makeroom]{cancel}
\usepackage{amssymb}
\usepackage{comment}
\usepackage{float}
\usepackage[dvipsnames]{xcolor}
\usepackage{ifpdf}
\usepackage{cite}
\usepackage{hyperref}
\usepackage{enumitem}
\usepackage{algorithm2e}
\usepackage{bbold}
\usepackage{gensymb}
\usepackage{multicol}

\begin{document}

\title{A Bayesian Approach to Forced Oscillation Source Location Given Uncertain Generator Parameters}

\author{Samuel Chevalier,~\IEEEmembership{Student Member,~IEEE,}
        Petr Vorobev,~\IEEEmembership{Member,~IEEE,}
        Konstantin Turitsyn,~\IEEEmembership{Member,~IEEE}
\thanks{This work was supported in part Skoltech-MIT Next Generation grant, and the MIT Energy Initiative Seed Fund Program.

S. Chevalier and K. Turitsyn are with Department of Mechanical Engineering, Massachusetts Institute of Technology. E-mail: schev, turitsyn@mit.edu

P. Vorobev is with Skolkovo Institute of Science and Technology and also with the Department of Mechanical Engineering, Massachusetts Institute of Technology. E-mail: petrvoro@mit.edu}}

\maketitle

\begin{abstract}
Since forced oscillations are exogenous to dynamic power system models, the models by themselves cannot predict when or where a forced oscillation will occur. Locating the sources of these oscillations, therefore, is a challenging problem which requires analytical methods capable of using real time power system data to trace an observed oscillation back to its source. The difficulty of this problem is exacerbated by the fact that the parameters associated with a given power system model can range from slightly uncertain to entirely unknown. In this paper, a Bayesian framework, via a two-stage Maximum A Posteriori optimization routine, is employed in order to locate the most probable source of a forced oscillation given an uncertain prior model. The approach leverages an equivalent circuit representation of the system in the frequency domain and employs a numerical procedure which makes the problem suitable for real time application. The derived framework lends itself to successful performance in the presence of PMU measurement noise, high generator parameter uncertainty, and multiple forced oscillations occurring simultaneously. The approach is tested on a 4-bus system with a single forced oscillation source and on the WECC 179-bus system with multiple oscillation sources.
\end{abstract}


\IEEEpeerreviewmaketitle


\section{Introduction}\label{Introduction}

\IEEEPARstart{W}{ith} the continued wide-scale deployment of Phasor Measurement Units (PMUs) across the transmission grid, power system operators have an increasingly adept ability to observe and respond to low frequency oscillations. Of particular concern are forced oscillations (FOs), which generally refer to a system's response to an external periodic disturbance~\cite{GHORBANIPARVAR:2017}. A broad range of causes~\cite{NERC:2017,GHORBANIPARVAR:2017}, such as control valve malfunctions, and resulting detrimental effects~\cite{Nezam:2016,Vanfretti:2012}, such as power quality degradation, are attributed to FOs.

Across industry and academia, there is general consensus that the most effective way to deal with an FO is to locate the component which is the source of the oscillation and disconnect it from service. Accordingly, a range of oscillation source location algorithms, many of which are outlined in~\cite{Wang:2017}, have recently been developed. Most notable is the Dissipating Energy Flow (DEF) method~\cite{MaslennikovPES:2017} which tracks the system-wide flow of so-called ``transient energy" in order to locate the source. This method has been successfully applied to many actual FO events in the ISO New England and WECC systems. Although successful in practice, a number of questions related to its modeling assumptions still remain open~\cite{Chen:2017,Chevalier:2018}. For example, in~\cite{Chevalier:2018}, it was shown that a constant impedance load may cause the DEF to locate the incorrect source of an FO. Since resistance in the system has been shown to act as an indefinite (positive or negative) energy source, other system elements which have not yet been considered may also act as indefinite energy sources, so further study is still needed before the DEF is universally applicable.

Non energy-based methods have also shown great promise. In reference~\cite{Cabrera:2017}, eigenvalue decomposition of the linearized system's state matrix is used in conjunction with the FO's measured characteristics to perform source location identification. The authors of~\cite{Meng:2017} employ machine learning techniques, via multivariate time series analysis, to perform source identification; all off-line classifier training is based on simulated data. A fully data driven method, which employs convex relaxation to optimally locate sparse FO sources, is introduced in~\cite{Huang:2018}. Due to the characteristically narrow bandwidth of FOs, other authors have embraced frequency domain techniques. In reference~\cite{Agrawal:2017}, the pseudo-inverse of a set of system transfer functions are multiplied by a vector of PMU measurements to yield an FO solution vector. Similarly,~\cite{Chevalier:2018} introduces a method for building the frequency response function (FRF) associated with a dynamic generator model. The FRF is then used to predict generator responses to terminal voltage oscillations: large deviation between the prediction and the PMU measurements at the forcing frequency indicate the source of an FO.

Model based source location algorithms incorporate the unfortunate drawback of solution accuracy being constrained by the accuracy of the model parameters used in the analysis. Purely data driven approaches, on the other hand, do not leverage known system structure and dynamics. To mitigate these drawbacks, this paper introduces a Bayesian approach for performing source identification, where PMU data is incorporated in a likelihood function and confidence in the system model is incorporated in a prior function. Others have applied Bayesian analysis to power systems in past. For example,~\cite{Bogodorova:2017} used a Bayesian particle filter for power plant parameter estimation, and~\cite{Petra:2017} solved a maximum a posteriori (MAP) optimization problem in the time domain to perform power system parameter identification. While parameter estimation is an important aspect of our solution to the FO source location problem, our primary goal is to locate the sources of the FOs.

Building off the FRF analysis presented in~\cite{Chevalier:2018}, the primary contributions of this paper are as follows.
\begin{enumerate}
\item A likelihood function and its physically meaningful covariance matrix are derived with respect to a generator's terminal signal perturbations in the frequency domain.
\item A Bayesian source location algorithm, via two-stage MAP optimization, is formulated to find the most likely set of dynamic model parameters and FO injection terms.
\item A numerical procedure is given which engenders computational tractability in the context of large scale systems.
\end{enumerate}

The remainder of this paper is structured as follows. In Section \ref{Problem Formulation}, we introduce how the MAP framework may be applied to the FO source location problem in the frequency domain. Section \ref{Location Algorithm} then provides an explicit, step-by-step algorithm for implementing the given procedure. Test results from a 4-bus system and the WECC 179-bus system of~\cite{Maslennikov:2016} are provided in Section \ref{Test Results}. Finally, concluding remarks are offered in Section \ref{Conclusion}.


\section{Problem Formulation}\label{Problem Formulation}
In this section, we first recall the concept of a generator's frequency response function (FRF) which was introduced in \cite{Chevalier:2018}. Next, we introduce the maximum a posteriori (MAP) framework --- the central concept of this manuscript --- and show how it may be leveraged for locating the sources of FOs. Finally, we present a MAP solution technique.

\subsection{Representing Generators as Admittance Matrices}\label{Admittance Representation}
\begin{figure}
\includegraphics[scale=0.66]{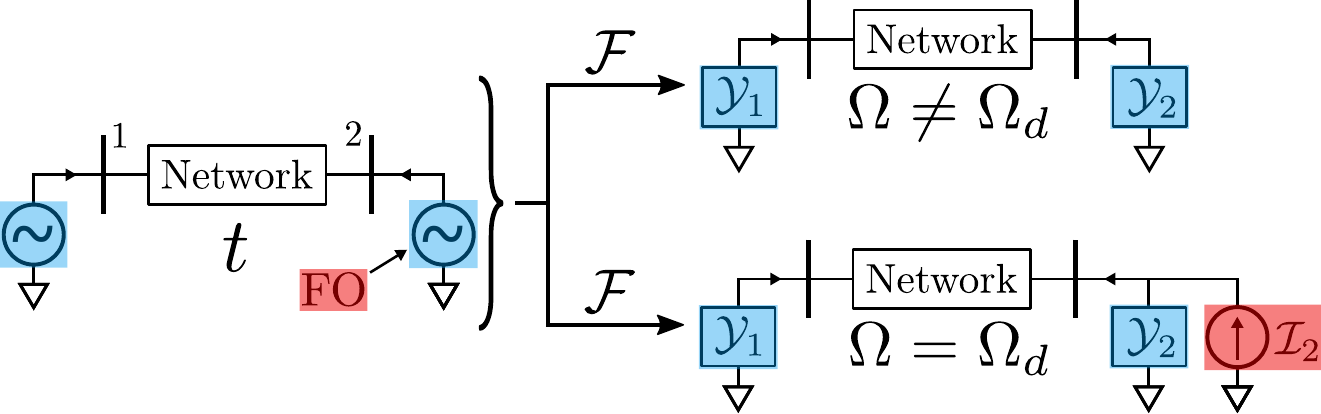}
\caption{\label{fig: Eq_Sys_Omegas} An FO in a time domain power system model (left) shows up as a current injection source at the forcing frequency $\Omega_d$ (bottom right) when the system is transformed to the frequency domain ($\mathcal F$). While generator 1 (a non-source bus) is transformed into just its FRF, generator 2 (a source bus) is transformed into its FRF plus a parallel current source when $\Omega=\Omega_d$. When $\Omega\ne\Omega_d$, this current injection is absent (top right).}
\end{figure}

As in \cite{Chevalier:2018}, we start by considering a generator\footnote{While this paper exclusively considers generators as the sources of FOs, the given framework may be extended to any dynamic system component.} connected to a power system and assuming PMU data from the generator's terminal bus are available. We also assume the generator is operating with steady state terminal voltage phasor ${\rm V}_{\!0}e^{j\theta_0}$ and current phasor ${\rm I}_0e^{j\phi_0}$. We respectively define ${\rm V}(t)$, $\theta(t)$, ${\rm I}(t)$ and ${\phi}(t)$ to be the \textit{measured} voltage magnitude, voltage phase, current magnitude and current phase \textit{deviations} from steady state. The Fourier transform $\mathcal F$ of these signals is
\begin{align}\label{fouriermain}
\tilde{X}(\Omega) & =\int_{\infty}^{-\infty}X(t)e^{j\Omega t}{\rm d}t,\quad X\in\{{\rm V},\;\theta,\;{\rm I},\;\phi\}.
\end{align}
To avoid confusion, we note that the frequency $\Omega$ has nothing to do with the fundamental AC frequency of $50$ or $60$ Hz; the transformation in \eqref{fouriermain} is performed on phasors with AC frequencies already excluded. We now consider the FRF $\mathcal{Y}\!\equiv\! \mathcal{Y}(\Omega)$ from \cite{Chevalier:2018} which relates a generator's terminal current (magnitude and phase) perturbations to its terminal voltage (magnitude and phase) perturbations in the frequency domain:
\begin{equation}\label{eq: Admittance}
\underbrace{\left[\begin{array}{c}
\tilde{{\rm I}}(\Omega)\\
\tilde{\phi}(\Omega)
\end{array}\right]}_\mathlarger{{\tilde {\bf I}}} \approx\underbrace{\left[\begin{array}{cc}
\mathcal{Y}_{11} & \mathcal{Y}_{12}\\
\mathcal{Y}_{21} & \mathcal{Y}_{22}
\end{array}\right]}_\mathlarger{{\mathcal Y}}\underbrace{\left[\begin{array}{c}
\tilde{{\rm V}}(\Omega)\\
\tilde{\theta}(\Omega)
\end{array}\right]}_\mathlarger{{\tilde {\bf V}}},\quad\Omega\ge0
\end{equation}
where the approximation is employed since measurement noise prevents (\ref{eq: Admittance}) from constituting an exact relationship. We refer to the vector $\tilde{\bf{I}}$ as the current \textit{measurement}, since it is directly measured by PMUs. Similarly, we refer to the vector $\mathcal{Y}\tilde{\bf{V}}$ as the current \textit{prediction}, since the measured voltage vector $\tilde{\bf{V}}$ is multiplied by an admittance matrix model to yield a current estimate. The error between the measured ($\tilde{{\bf I}}$) and predicted ($\mathcal{Y}\tilde{{\bf V}}$) currents, at each frequency $\Omega$, may be quantified via the $\ell_2$ norm:
\begin{align}\label{eq: l2_error}
\text{prediction error}=\left\Vert \tilde{{\bf I}}-\mathcal{Y}\tilde{{\bf V}}\right\Vert_2.
\end{align}
For small perturbations, the primary contributors to the prediction error are PMU measurement noise, generator model parameter inaccuracies, and unmodeled generator inputs such as external perturbations. It is further shown in~\cite{Chevalier:2018} that when generators are transformed into their equivalent FRFs, any FO at a source bus acts as a current injection represented by the $2\times1$ complex vector $\mathcal{I}$, where $\mathcal{I}=[\mathcal{I}_{\rm I}\;\mathcal{I}_{\phi}]^{\top}$ is used to denote the complex current magnitude ($\mathcal{I}_{\rm I}$) and complex current phase ($\mathcal{I}_{\phi}$) injections around particular forcing frequency $\Omega_d$:
\begin{align}\label{eq: I_inj}
\tilde{{\bf I}}&\approx\mathcal{Y}\tilde{{\bf V}}+\mathcal{I}.
\end{align}
This relation is illustrated at Fig. \ref{fig: Eq_Sys_Omegas}.
Since FOs are usually dominant at some forcing frequency $\Omega_d$ and nonexistent elsewhere in the frequency spectrum  (neglecting nonlinear harmonics), the current injection ${\mathcal I}$ will be equal to $\bf 0$ for all frequencies $\Omega \ne\Omega_d$. At any generator which is not the FO source, the current injection is zero for all $\Omega$. 

\begin{figure}
\includegraphics[scale=0.495]{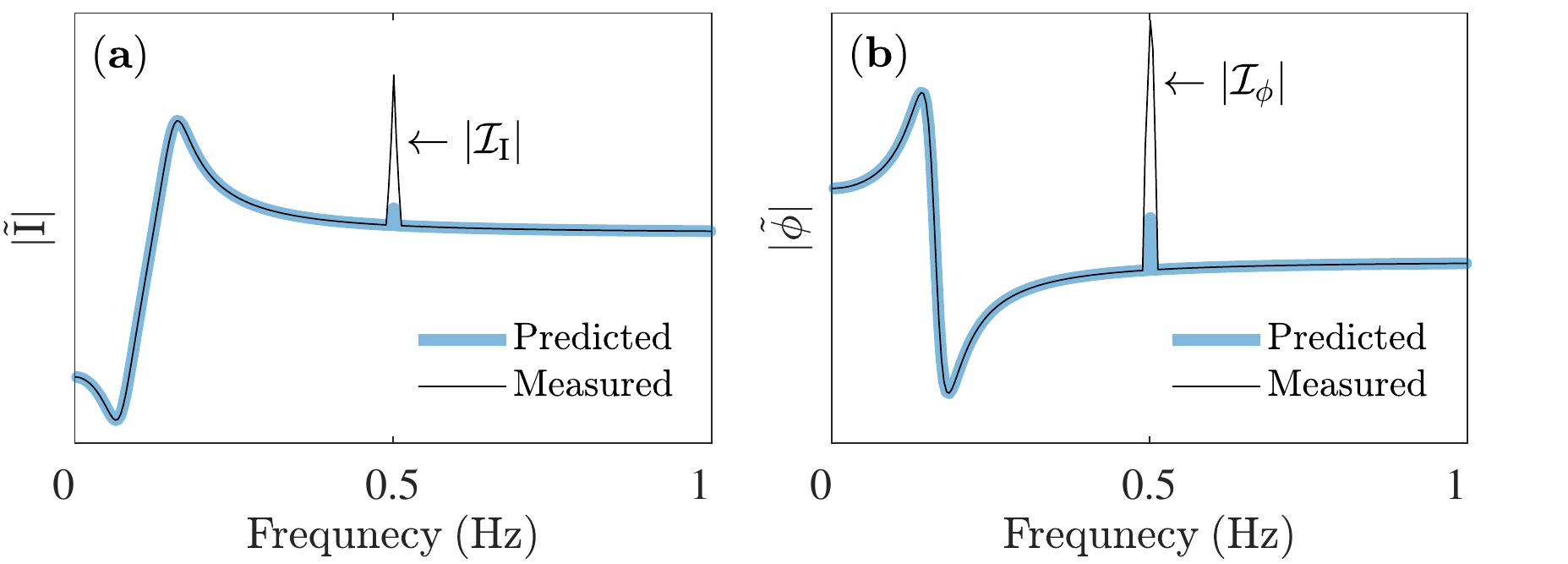}
\caption{\label{fig: I_inj_Example} The modulus of the measured and predicted current magnitude $\tilde{\rm I}$ and current phase $\tilde{\phi}$ spectrums at a source generator are plotted in panels ($\bf a$) and ($\bf b$), respectively. At the forcing frequency of $f_{d}=0.5$ Hz, the current injection magnitudes $|\mathcal{I}_{\rm I}|$ and $|\mathcal{I}_{\phi}|$ can be seen as deviations from the predicted terminal current spectrum.}
\end{figure}

An example of the measured and predicted current spectrums associated with a source generator (in the absence of measurement noise and model uncertainty) is shown by Fig. \ref{fig: I_inj_Example}. Panels ($\bf a$) and ($\bf b$) show the presence of the FO current injection magnitudes which cause deviation between the measured and predicted current spectrums. At non-source generators, these injections do not exist and the measurements and predictions match.

\subsection{Constructing the System Likelihood Function}\label{Likelihood}

Identifying the source of an FO based on a generator's FRF, as outlined in the previous subsection, relies on the knowledge of the generator model in order to construct its $\mathcal{Y}$-matrix. If the generator model parameters are not known with sufficient accuracy, the method cannot be applied directly. However, since the current injection function $\mathcal{I}$ is only non-zero in a narrow band around the forcing frequency $\Omega_d$, one can use the generator's measured response in the remainder of the spectrum in order to identify its parameters. This is accomplished by employing the MAP framework. 

To construct the likelihood function which characterizes the admittance matrix relationship of (\ref{eq: I_inj}), let us first consider the following generalized dynamical system: 
\begin{equation}\label{eq: y=Ax}
{\bf y}=A{\bf x}+\boldsymbol{\eta}
\end{equation}
where $\boldsymbol{\eta}$ is some independent identically distributed (IID) vector of additive white Gaussian noise (AWGN) variables with $\eta_i \sim {\mathcal N}(0,\sigma^2)$ and $\bf x$ and $\bf y$ are input and output vectors. The covariance matrix of the likelihood function is thus
\begin{align}
{\rm E}\left[\left({\bf y}-A {\bf x}\right)\left({\bf y}-A {\bf x}\right)^{\top}\right] & ={\rm E}\left[\boldsymbol{\eta}\boldsymbol{\eta}^{\top}\right].
\end{align}
Since $\boldsymbol{\eta}$ is an IID vector, then $\Gamma_{\boldsymbol{\eta}}\coloneqq{\rm E}[\boldsymbol{\eta}\boldsymbol{\eta}^{\top}] = \sigma^{2}\mathbb{1}$ where $\mathbb{1}$ is the identity matrix. The multivariate normal distribution associated with this system is thus given by 
\begin{align}
p_{\boldsymbol{\eta}}(\eta_{1},...,\eta_{r}) & =\frac{e^{-\frac{1}{2}\left({\bf y}-A{\bf x}\right)^{\top}\Gamma_{\boldsymbol{\eta}}^{-1}\left({\bf y}-A{\bf x}\right)}}{\sqrt{\left(2\pi\right)^{l}{\rm det}\left(\Gamma_{\boldsymbol{\eta}}\right)}}.\label{eq: p_simple}
\end{align}
Since the measurement noise associated with PMU data is approximately white~\cite{Brown:2016}, a similar likelihood function may be constructed for the system in (\ref{eq: I_inj}), where ``system" in this subsection refers to a single generator. To build this likelihood function, the expanded right hand side (RHS) of (\ref{eq: I_inj}) is subtracted from the expanded left hand side (LHS). Assuming an accurate admittance matrix, the system dynamics cancels out and only measurement noise terms are left over on the RHS of (\ref{eq: Diff_Noise}):
\begin{align}
\left[\begin{array}{c}\nonumber
\tilde{{\rm I}}\\
\tilde{\phi}
\end{array}\right]-\left[\begin{array}{cc}
\mathcal{Y}_{11} & \mathcal{Y}_{12}\\
\mathcal{Y}_{21} & \mathcal{Y}_{22}
\end{array}\right]\left[\begin{array}{c}
\tilde{{\rm V}}\\
\tilde{\theta}
\end{array}\right]&-\left[\begin{array}{c}
\mathcal{I}_{{\rm I}}\\
\mathcal{I}_{\phi}
\end{array}\right]=\\ 
\left[\!\begin{array}{c}
\tilde{\epsilon}_{\rm I}\\
\tilde{\epsilon}_{\phi}
\end{array}\!\right]-\left[\begin{array}{cc}
\mathcal{Y}_{11} & \mathcal{Y}_{12}\\
\mathcal{Y}_{21} & \mathcal{Y}_{22}
\end{array}\right]\left[\!\begin{array}{c}
\tilde{\epsilon}_{\rm V}\\
\tilde{\epsilon}_{\theta}
\end{array}\!\right]&\label{eq: Diff_Noise}
\end{align}
where $\tilde{\epsilon}_{X}$, $X\!\in\!\{{\rm V},\,\theta,\,{\rm I},\,\phi\}$ is a complex random variable associated with the frequency domain representation of the AWGN distribution $\epsilon_X$. Accordingly, measured signal $X(t)$ and its associated true signal $\overline{X}(t)$ are related by
\begin{align}\label{eq: meas_true}
X(t) = \overline{X}(t)+\epsilon_X,\;\;\; X\!\in\!\{{\rm V},\,\theta,\,{\rm I},\,\phi\}.
\end{align}
We now assume $\epsilon_X$ is sampled $2K\!+\!1$ times with sample rate $f_s$. The values are placed into discrete vector $\epsilon_X[n]$, and its Discrete Fourier Transform\footnote{We define the $2K\!+\!1$ point double sided DFT of $x[n]$ as $\tilde{x}[w]=\sum_{k=0}^{2K}x[n]e^{-j\frac{2\pi wn}{2K+1}}$ where $w=0,1,...,2K$.} (DFT) is taken. The resulting single-sided output, ${\tilde \epsilon}_X[w]$, will be a function of the $K\!+\!1$ frequencies
\begin{align}\label{eq: Omega_w}
\Omega_w=2\pi\times\left[0,\;\frac{f_{s}}{2K\!+\!1},\;\frac{2\cdot f_{s}}{2K\!+\!1},\ldots,\frac{K \cdot f_{s}}{2K\!+\!1}\right].
\end{align}
Since the input distribution to the DFT is a Gaussian, the output will be a frequency dependent set of complex Gaussians which are IID across frequency. Accordingly, ${\tilde \epsilon}_X$ in fact does not depend on frequency since ${\tilde{\epsilon}}_{X}[w_1]$ and ${\tilde{\epsilon}}_{X}[w_2]$ are distributed identically. Additionally, ${\tilde \epsilon}_X$ can be split into its real and imaginary components via ${\tilde{\epsilon}_{X}}={\tilde{\epsilon}_{X_{r}}+j{\tilde{\epsilon}_{X_{i}}}}$, where $\tilde{\epsilon}_{X_{r}}$ and $\tilde{\epsilon}_{X_{i}}$ are both real valued, IID Gaussians with $\rm E[\tilde{\epsilon}_{X_{r}}]=\rm E[\tilde{\epsilon}_{X_{i}}]=0$. By the central limit theorem and basic statistics, the variances of $\tilde{\epsilon}_{X_{r}}$ and $\tilde{\epsilon}_{X_{i}}$, which are essential for eventually building the likelihood covariance matrix, can be computed:
\begin{align}
\rm E[\tilde{\epsilon}_{X_{r}}^2]=\rm E[\tilde{\epsilon}_{X_{i}}^2]=\frac{4\left(2K\!+\!1\right)}{2}{\rm E}[\epsilon^2_X].\label{eq: PMU_fft_noise}
\end{align}
To continue building the likelihood function, (\ref{eq: Diff_Noise}) must be separated into its real and imaginary parts in order to preserve the Gaussian nature of the measurement noise. It is important to note that there are no measurement noise terms associated with the current injections since vector $\mathcal I$ is not measured but is instead a mathematical artifact which represents the current flow attributed to an FO. For notational convenience, the LHS of (\ref{eq: Diff_Noise}) may be rewritten in terms of the complex variables ${\tilde{M}}=\tilde{M}_r+j\tilde{M}_i$ (magnitude) and ${\tilde{P}}=\tilde{P}_r+j\tilde{P}_i$ (phase) while the RHS of (\ref{eq: Diff_Noise}) may be rewritten in terms of the corresponding complex noise variables ${\tilde{N}}=\tilde{N}_r+j\tilde{N}_i$ and ${\tilde{Q}}=\tilde{Q}_r+j\tilde{Q}_i$:
\begin{align}
\left[\begin{array}{c}\label{eq: RI_Admittance}
\tilde{M}_{r}+j\tilde{M}_{i}\\
\tilde{P}_{r}+j\tilde{P}_{i}
\end{array}\right]=\left[\begin{array}{c}
\tilde{N}_{r}+j\tilde{N}_{i}\\
\tilde{Q}_{r}+j\tilde{Q}_{i}
\end{array}\right].
\end{align}
Equation (\ref{eq: RI_Admittance}) is valid across all frequencies, and it is entirely analogous to writing (\ref{eq: y=Ax}) as ${\bf y}-A{\bf x}={\boldsymbol \eta}$. Explicitly, the residual expressions on the LHS of (\ref{eq: RI_Admittance}) take on the following forms:
\begin{align}
\tilde{M}_{r} & \!=\tilde{{\rm I}}_{r}-\mathcal{Y}_{11r}\tilde{{\rm V}}_{r}+\mathcal{Y}_{11i}\tilde{{\rm V}}_{i}-\mathcal{Y}_{12r}\tilde{\theta}_{r}+\mathcal{Y}_{12i}\tilde{\theta}_{i}\!-\!{\mathcal I}_{{\rm I}_r}\label{eq: Mr}\\
\tilde{M}_{i} & \!=\tilde{{\rm I}}_{i}-\mathcal{Y}_{11i}\tilde{{\rm V}}_{r}\!-\!\mathcal{Y}_{11r}\tilde{{\rm V}}_{i}-\mathcal{Y}_{12i}\tilde{\theta}_{r}-\mathcal{Y}_{12r}\tilde{\theta}_{i}-{\mathcal I}_{{\rm I}_i}\label{eq: Mi}\\
\tilde{P}_{r} & \!=\tilde{\phi}_{r}-\mathcal{Y}_{21r}\tilde{{\rm V}}_{r}+\mathcal{Y}_{21i}\tilde{{\rm V}}_{i}-\mathcal{Y}_{22r}\tilde{\theta}_{r}+\mathcal{Y}_{22i}\tilde{\theta}_{i}\!-\!{\mathcal I}_{{\phi}_r}\label{eq: Pr}\\
\tilde{P}_{i} & \!=\tilde{\phi}_{i}-\mathcal{Y}_{21i}\tilde{{\rm V}}_{r}-\mathcal{Y}_{21r}\tilde{{\rm V}}_{i}-\mathcal{Y}_{22i}\tilde{\theta}_{r}-\mathcal{Y}_{22r}\tilde{\theta}_{i}\!-\!{\mathcal I}_{{\phi}_i}\label{eq: Pi}
\end{align}
where the subscripts $r$ and $i$ denote the real and imaginary parts of the admittance matrix entries, complex frequency domain signals, and current injection terms. The noise-related expressions, on the RHS of (\ref{eq: RI_Admittance}), take on the following forms:
\begin{align}
\tilde{N}_{r} & =\tilde{\epsilon}_{{\rm I}_{r}}-\mathcal{Y}_{11r}\tilde{\epsilon}_{{\rm V}_{r}}+\mathcal{Y}_{11i}\tilde{\epsilon}_{{\rm V}_{i}}-\mathcal{Y}_{12r}\tilde{\epsilon}_{\theta_{r}}+\mathcal{Y}_{12i}\tilde{\epsilon}_{\theta_{i}}\label{eq: Nr}\\
\tilde{N}_{i} & =\tilde{\epsilon}_{{\rm I}_{i}}-\mathcal{Y}_{11i}\tilde{\epsilon}_{{\rm V}_{r}}-\mathcal{Y}_{11r}\tilde{\epsilon}_{{\rm V}_{i}}-\mathcal{Y}_{12i}\tilde{\epsilon}_{\theta_{r}}-\mathcal{Y}_{12r}\tilde{\epsilon}_{\theta_{i}}\\
\tilde{Q}_{r} & =\tilde{\epsilon}_{\phi_{r}}-\mathcal{Y}_{21r}\tilde{\epsilon}_{{\rm V}_{r}}+\mathcal{Y}_{21i}\tilde{\epsilon}_{{\rm V}_{i}}-\mathcal{Y}_{22r}\tilde{\epsilon}_{\theta_{r}}+\mathcal{Y}_{22i}\tilde{\epsilon}_{\theta_{i}}\\
\tilde{Q}_{i} & =\tilde{\epsilon}_{\phi_{i}}-\mathcal{Y}_{21i}\tilde{\epsilon}_{{\rm V}_{r}}-\mathcal{Y}_{21r}\tilde{\epsilon}_{{\rm V}_{i}}-\mathcal{Y}_{22i}\tilde{\epsilon}_{\theta_{r}}-\mathcal{Y}_{22r}\tilde{\epsilon}_{\theta_{i}}\label{eq: Qi}
\end{align}
where the real and imaginary components of the measurement noise distributions, as characterized by (\ref{eq: PMU_fft_noise}), have been employed explicitly. In equating the real and imaginary parts of (\ref{eq: RI_Admittance}), four equations are yielded: $\tilde{M}_{r}=\tilde{N}_{r}$, $\tilde{M}_{i}=\tilde{N}_{i}$, $\tilde{P}_{r}=\tilde{Q}_{r}$, and $\tilde{P}_{i}=\tilde{Q}_{i}$. Assuming each of these can be written for the $K+1$ frequencies of (\ref{eq: Omega_w}), then (\ref{eq: RI_Admittance}) may be used to generate a total of $4(K+1)$ equations. It is thus useful to define the following $\mathbb{R}^{K+1}$ function vectors:
\begin{align}
\begin{array}{cc}
{\bf Z}_{r} & \!\!\!\!\!\coloneqq\left[\tilde{Z}_{r}[0],\ldots,\tilde{Z}_{r}[K]\right]^{\top}\\
{\bf Z}_{i} & \!\!\!\!\!\!\coloneqq\left[\tilde{Z}_{i}[0],\ldots,\tilde{Z}_{i}[K]\right]^{\top}
\end{array}\!\!\!, \,\;Z\!\in\{M,N,P,Q\}
\end{align}
where each vector entry is a function of one of the $K+1$ frequencies from vector (\ref{eq: Omega_w}). The full noise function vector $\bf{L}$ is now defined which represents the concatenation of the four individual noise function vectors:
\begin{align}
{\bf L}=[{\bf N}_{r}^{\top}\;{\bf N}_{i}^{\top}\;{\bf Q}_{r}^{\top}\;{\bf Q}_{i}^{\top}]^{\top}.\label{eq: L}
\end{align}
The residual function vector $\bf{R}$ may be defined similarly:
\begin{align}
{\bf R}=[{\bf M}_{r}^{\top}\;{\bf M}_{i}^{\top}\;{\bf P}_{r}^{\top}\;{\bf P}_{i}^{\top}]^{\top}.\label{eq: R}
\end{align}
The $4(K\!+\!1)\!\times\!4(K\!+\!1)$ covariance matrix of ${\bf L}$ is thus
\begin{align}
\Gamma_{{\bf L}} & ={\rm E}\left[{\bf L}{\bf L}^{\top}\right]\label{Gamma_{L}}\\
 & =\left[\begin{array}{cccc}
\Gamma_{{\bf N}_{r}} & {\bf 0} & \Gamma_{{\bf N}_{r}{\bf Q}_{r}} & \Gamma_{{\bf N}_{r}{\bf Q}_{i}}\\
{\bf 0} & \Gamma_{{\bf N}_{i}} & \Gamma_{{\bf N}_{i}{\bf Q}_{r}} & \Gamma_{{\bf N}_{i}{\bf Q}_{i}}\\
\Gamma_{{\bf Q}_{r}{\bf N}_{r}} & \Gamma_{{\bf Q}_{r}{\bf N}_{i}} & \Gamma_{{\bf Q}_{r}} & {\bf 0}\\
\Gamma_{{\bf Q}_{i}{\bf N}_{r}} & \Gamma_{{\bf Q}_{i}{\bf N}_{i}} & {\bf 0} & \Gamma_{{\bf Q}_{i}}
\end{array}\right]\label{Gamma2_{L}}
\end{align}
where the zero matrices $\bf 0$ are inserted due to the fact that ${\rm E}[{\tilde N}_r{\tilde N}_i]={\rm E}[{\tilde Q}_r{\tilde Q}_i]=0$, by inspection, for all frequencies. By direct extension, $\Gamma_{{\bf N}_{r}{\bf N}_{i}}=\Gamma_{{\bf Q}_{r}{\bf Q}_{i}}={\bf 0}$. Each of the non-zero sub-covariance matrices in (\ref{Gamma2_{L}}) will be diagonal due to the fact that the frequency domain representation of measurement noise at the $k^{\rm th}$ frequency is uncorrelated with everything except for itself at the particular $k^{\rm th}$ frequency. 

The multivariate Gaussian likelihood function may now be constructed. In Bayesian analysis, the likelihood probability density function (PDF) $p_{{\rm likely}}$ quantifies the likelihood of the observed data in $\bf d$ given some set of model parameters in $\Theta$. In this context, $\Theta$ is a vector filled with the generator parameters $\Theta_{g}$ which are necessary to construct $\mathcal Y$ (such as reactances, time constants or AVR gains) and the current injection terms $\Theta_{\mathcal{I}}$:
\begin{equation}\label{eq: Theta_full}
\Theta=\left\{ \begin{array}{cl}
\Theta_{g} & \Rightarrow \text{ generator parameters}\\
\Theta_{\mathcal{I}} & \Rightarrow \text{ current injection terms}.
\end{array}\right.
\end{equation}
Entirely analogous to (\ref{eq: p_simple}), the likelihood function itself is
\begin{align}
p_{\rm likely}({\bf d}|\Theta) & =\frac{e^{-\frac{1}{2}{\bf R}^{\top}\Gamma_{{\bf L}}^{-1}{\bf R}}}{\sqrt{\left(2\pi\right)^{4(K\!+\!1)}{\rm det}\left(\Gamma_{{\bf L}}\right)}}.\label{eq: likelyT}
\end{align}

\subsection{Constructing the System Prior Function}\label{Priors}
\begin{figure}
\begin{centering}
\includegraphics[scale=0.85]{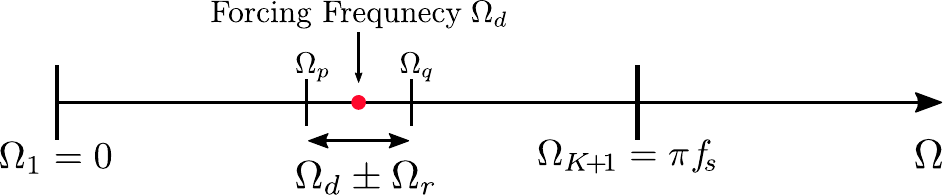}
\caption{\label{fig: Freq_Range}The frequencies between $\Omega_p=\Omega_d-\Omega_r$ and $\Omega_q=\Omega_d+\Omega_r$ represent the frequency range over which the FO has significant effect on the system. $\Omega_r$ is half the range over which the forcing frequency has effect.}
\end{centering}
\end{figure}

Typically, the generator model parameter values in the vector $\Theta_g\in\mathbb{R}^{m}$ are not certain, and it is common to quantify this initial certainty with another multivariate Gaussian PDF~\cite{Petra:2017}:
\begin{align}\label{eq: prior_gen}
p_{{\rm prior1}}(\Theta_{g}) & =\frac{e^{-\frac{1}{2}\left(\Theta_{g}-\overline{\Theta}_{g}\right)^{\top}\Gamma_{g}^{-1}\left(\Theta_{g}-\overline{\Theta}_{g}\right)}}{\sqrt{\left(2\pi\right)^{m}{\rm det}\left(\Gamma_{g}\right)}}
\end{align}
where $\overline{\Theta}_{g}$ is the mean vector of prior generator parameter constants and $\Gamma_{g}$ is the corresponding diagonal covariance matrix. High model parameter confidence corresponds to low variance values. Also contained in $\Theta$ is the vector $\Theta_{\mathcal{I}}\in\mathbb{R}^{4v}$:
\begin{align}\label{eq: I_concat}
\Theta_{\mathcal{I}}=\left\{ \begin{array}{c}
\mathcal{I}_{{\rm I}_{r}}(\Omega)\\
\mathcal{I}_{{\rm I}_{i}}(\Omega)\\
\mathcal{I}_{\phi_{r}}(\Omega)\\
\mathcal{I}_{\phi_{i}}(\Omega)
\end{array}\!\!,\;\Omega\in\{\Omega_{p}\ldots\Omega_{q}\}\right.
\end{align}
where $\mathcal{I}_{{\rm I}_{r}}$, $\mathcal{I}_{{\rm I}_{i}}$, $\mathcal{I}_{\phi_{r}}$ and $\mathcal{I}_{\phi_{i}}$ are used in (\ref{eq: Mr})-(\ref{eq: Pi}). By leveraging prior knowledge about the central FO frequency $\Omega_d$, we define these injections to exist only across the small range of DFT frequencies where the FO energy is dominant: $\Omega_p$ through $\Omega_q$ as shown by Fig. \ref{fig: Freq_Range}. In defining $v$ discrete frequencies is this range, there are a total of $4v$ current injection parameters (per FO) at each generator to include in (\ref{eq: I_concat}).
The prior distribution for these parameters will be taken as an unconditional IID Laplace distribution~\cite{Park:2008} (this choice shall be justified at the end of subsection \ref{Apply_MAP}):
\begin{align}\label{eq: prior_current}
p_{\rm prior2}(\Theta_{\mathcal{I}})=\prod_{j=1}^{4v}\frac{\lambda}{2}e^{-\lambda\left|\Theta_{\mathcal{I}_{j}}\right|}
\end{align}
where $\Theta_{\mathcal{I}_{j}}$ refers to the $j^{\rm th}$ current injection variable from (\ref{eq: I_concat}). The generator's full prior is the product of the generator parameter prior (\ref{eq: prior_gen}) and the current injection prior (\ref{eq: prior_current}).

\subsection{Applying the MAP Formulation to the Forced Oscillation Source Location Problem}\label{Apply_MAP}
By leveraging the given likelihood and prior functions, a Bayesian framework, via MAP optimization, can be used to locate the sources of FOs in the context of a power system with poorly known generator parameters. The posterior distribution, which represents the likelihood of the model parameters given the data that have been observed, is computed through the application of Bayes' rule at each generator:
\begin{align}\label{eq: post}
p_{{\rm post}}(\Theta|{\bf d}) \propto p_{{\rm likely}}({\bf d}|\Theta)p_{{\rm prior1}}(\Theta_g)p_{{\rm prior2}}(\Theta_{\mathcal I}).
\end{align}
We now seek to maximize the posterior since ${\rm max}\{{p_{\rm post}}\}$ corresponds to maximum confidence in the model parameters for a given set of observed data. Maximizing this distribution is equivalent to minimizing the negative of its natural log\footnote{To make this statement, we assume the determinant of the covariance matrix in (\ref{eq: likelyT}) is roughly constant across plausible model parameters.}:
%
%
\begin{align}\label{eq: Theta_MAP}
\!\!\Theta_{{\rm MAP}}\! & =\underset{\Theta\in\mathbb{R}^{z}}{{\rm argmin}}\left\{ -\log(p_{{\rm post}}(\Theta|{\bf d}))\right\} \\
 & =\underset{\Theta\in\mathbb{R}^{z}}{{\rm argmin}}\left\{ \!\left\Vert \Theta_{g}\!\!-\!\overline{\Theta}_{g}\right\Vert _{\Gamma_{g}^{-1}}^{2}\!\!+\!{\bf R}^{\!\top}\Gamma_{{\bf L}}^{-1}{\bf R}\!+\!\!\lambda\left\Vert \Theta_{\mathcal{I}}\right\Vert _{1}\!\right\} 
\end{align}
where $||{\bf x}||_1=\sum_i|x_i|$ and $z=m+4v$. The unconstrained optimization problem formulated by (\ref{eq: Theta_MAP}) is similar in structure to the LASSO problem~\cite{Cai:2013} with the regularization parameter $\lambda$ acting as a penalty on non-sparse solutions for the vector $\Theta_{\mathcal{I}}$, although the admittance matrix $\mathcal Y$ contained in $\bf R$ is highly nonlinear. Generally, optimization problems with objective functions which are formulated as 
\begin{align}\label{eq: L_lim}
\min_{{\bf x},{\bf y}}	\;\;\;L({\bf x},{\bf y})+\lambda\left\Vert {\bf y}\right\Vert _{1}
\end{align}
are non-differentiable when elements of ${\bf y}$ are zero. A workaround for solving ($\ref{eq: L_lim}$) transforms the unconstrained problem into a constrained problem~\cite{Schmidt:2009} by introducing slack variables in the vector $\bf u$ such that (\ref{eq: L_lim}) may be restated as
\begin{equation}
\label{eq: l1_const}
\begin{aligned}
\min_{{\bf x},{\bf y},{\bf u}}	\;\;\;& L({\bf x},{\bf y})+\lambda\sum_{i}u_i\\
\text{s. t.}	\;\;\;& -{\bf u}\le{\bf y}\le{\bf u}.
\end{aligned}
\end{equation}
The unconstrained FO optimization problem may be restated similarly, where slack variable vector $\bf s$ has been introduced:
\begin{equation}
\label{eq: min_final}
\begin{aligned}
\text{Minimize}\quad & \left\Vert \Theta_{g}-\overline{\Theta}_{g}\right\Vert _{\Gamma_{g}^{-1}}^{2}+{\bf R}^{\top}\Gamma_{{\bf L}}^{-1}{\bf R}+\lambda\sum_{i=1}^{4v}s_{i}\\
\text{subject to\quad} & -{\bf s}\le\Theta_{\mathcal{I}}\le{\bf s}.
\end{aligned}
\end{equation}
This optimization problem is written for a single generator. We now consider a power system with $g$ generators of which one, or more, may be the source(s) of the FO(s); PMU data from each generator is assumed to be available. In defining the scalar cost function $f_i=\left\Vert \Theta_{g}-\overline{\Theta}_{g}\right\Vert _{\Gamma_{g}^{-1}}^{2}+{\bf R}^{\top}\Gamma_{{\bf L}}^{-1}{\bf R}+\lambda{\bf 1}^\top{\bf s}$ associated with the $i^{\rm th}$ generator, we may minimize the sum of these cost functions over $g$ generators:
\begin{equation}
\label{eq: min_sum}
\begin{aligned}
\text{Minimize}\quad   & \sum_{i=1}^{g}f_{i}\\
\text{subject to}\quad & -{\bf s}_{i}\le\Theta_{\mathcal{I}i}\le{\bf s}_{i},\;\;i\in\{1 \ldots g\}.
\end{aligned}
\end{equation}
Particularly useful is that (\ref{eq: min_sum}) may be solved as a set of $g$ \textit{uncoupled} optimization problems: one for each generator. This is made possible due to a relaxation introduced in choosing the current injection prior of (\ref{eq: prior_current}), which is now explained. If a system operator knows that an FO is occurring in a system, but the source generator is unknown, then the \textit{most appropriate} prior distribution for (\ref{eq: I_concat}) would be one which introduces an $\ell_0$ norm constraint on current injections among all system generators; this would constrain the number of non-zero current injections found in the system to be equal to the number of occurring FOs. Aside from the NP-harness associated with such a formulation~\cite{Cai:2013}, an $\ell_0$ norm constraint would require that the generator posterior distributions be optimized simultaneously, thus coupling the optimization problems of (\ref{eq: min_sum}) and introducing large computational burden. As a relaxed alternative, a Laplace prior~\cite{Park:2008} is chosen to quantify the initial confidence in the current injection parameters because it ultimately introduces an $\ell_1$ norm penalty in (\ref{eq: Theta_MAP}). This $\ell_1$ norm penalty naturally encourages sparse regression parameter selection~\cite{Park:2008} and introduces the benefit of uncoupled optimization despite inheriting the drawback of relaxed sparsity. As evidenced by LASSO's popularity, this is a common relaxation approach~\cite{Cai:2013} applied to problems which seek sparse parameter recovery.

%
%

\subsection{A Numerical Procedure for MAP Solution}\label{NH for MAP_Solution}
In numerically solving (\ref{eq: min_sum}), the problem becomes computationally burdensome if the true likelihood covariance matrix of (\ref{Gamma2_{L}}) is used in the objective function. Since $\Gamma_{\bf L}$ depends directly on the model parameters in $\mathcal Y$, obtaining an analytical solution for $\Gamma_{\bf L}^{-1}$, such that it can be used in computing the necessary gradients and Hessians, is computationally intensive. In order to minimize (\ref{eq: min_final}), the following heuristic steps effectively balance formulation fidelity with tractability:

\begin{enumerate}

\item At the $i^{\rm th}$ iterative optimization step, $\Gamma_{\bf L}$ is numerically evaluated with the $i^{\rm th}$ parameter values of $\Theta_g$ such that $\overline{\Gamma}_{{\bf L}_i}\equiv\left.\Gamma_{\bf L}\right|_{\Theta_{g_i}}$

\item Constant matrix $\overline{\Gamma}_{{\bf L}_i}$ replaces analytical matrix ${\Gamma}_{\bf L}$ in (\ref{eq: min_final})

\item One iterative step is taken in the direction which minimizes this altered cost function, thus computing $\Theta_{g_{i+1}}$

\item In returning to step 1, these steps are repeated for the $(i\!+\!1)^{\rm th}$ iteration, and so on

\end{enumerate}

This process is applied for the covariance matrices of all $g$ generators in (\ref{eq: min_sum}). In treating these matrices as numerically constant at each optimization step, 
the objective function for a single $i^{\rm th}$ generator may now be restated as
\begin{equation}
f_i=\left\Vert \Theta_{g}-\overline{\Theta}_{g}\right\Vert _{{\Gamma}_{g}^{-1}}^{2}+{\bf R}^{\top}\overline{\Gamma}_{{\bf L}}^{-1}{\bf R}+\lambda{\bf 1}^\top{\bf s}.\label{eq: obj_func_constG}
\end{equation}
Since $\frac{\rm d}{{\rm d}\Theta}(\overline{\Gamma}_{{\bf L}}^{-1})={\bf 0}$, Hessian element $i,j$ associated with $C\coloneqq{\bf R}^{\top}\overline{\Gamma}_{{\bf L}}^{-1}{\bf R}$ is therefore simply
\begin{align}
\frac{{\rm d}^{2}C}{{\rm d}\Theta_{j}{\rm d}\Theta_{i}} & =2\left(\frac{{\rm d}^{2}{\bf R}^{\top}}{{\rm d}\Theta_{j}{\rm d}\Theta_{i}}\overline{\Gamma}_{{\bf L}}^{-1}{\bf R}+\frac{{\rm d}{\bf R}^{\top}}{{\rm d}\Theta_{j}}\overline{\Gamma}_{{\bf L}}^{-1}\frac{{\rm d}{\bf R}}{{\rm d}\Theta_{i}}\right).
\end{align}
An interior point method may be used to solve the optimization problem set up by (\ref{eq: min_sum}) with objective function (\ref{eq: obj_func_constG}). 


%
%


\section{Defining a Forced Oscillation Source Location Algorithm}\label{Location Algorithm}
In this section, the formulations introduced in~\cite{Chevalier:2018} and Section \ref{Problem Formulation} are tied together to explicitly define an FO source location algorithm. For enhanced effectiveness, a two-stage Bayesian update optimization scheme is introduced. This two-stage scheme allows the optimizer to primarily focus on generator parameter selection in stage 1 (by excluding data in the bandwidth of the FO) and current injection selection in stage 2 (by tightening the variances of the generator parameters based on the results of stage 1).

\subsection*{Stage 1}
In stage 1, current injections are \textit{not} considered. This is made possible if the system operator has prior knowledge about the location, in the frequency spectrum, of the current injections (FOs). By instructing the optimizer to ignore current injection variables $\Theta_{\mathcal I}$ and all data in the FO range of Fig. \ref{fig: Freq_Range}, the optimizer is able to tune generator parameters without considering current injections. Since this optimization formulation does not incorporate injection variables, it is thus unconstrained and has the following form across $g$ generators:
\begin{equation}
\label{eq: stage1_eqs}
\begin{aligned}
\text{Minimize}\quad\! & \sum_{i=1}^{g}f_{i}\\
 & f_{i}=\left\Vert \Theta_{g}-\overline{\Theta}_{g}\right\Vert _{\Gamma_{g}^{-1}}^{2}+{\bf R}^{\top}\overline{\Gamma}_{{\bf L}}^{-1}{\bf R}.
\end{aligned}
\end{equation}

In populating the residual function vector $\bf R$ with PMU data, an important practical consideration which should be accounted for is the time window associated with the data. Since the methods of this paper are based on linear analysis, the FRF is a direct function of the equilibrium of the system. If this equilibrium shifts significantly, parameter estimation and current injection determination will not be possible. To minimize these nonlinear affects, a short time window (on the order of a few minutes) should be employed to ensure that the analysis is unaffected by equilibrium swings.

Once the optimizer has converged to some local minimum (termed $\Theta_{{\rm MAP}1}$) and stage 1 is complete, the resulting posterior distribution will have mean $\Theta_{{\rm MAP}1}$ and covariance matrix ${\Gamma}_{\Theta_{{\rm MAP}1}}$ equal to the inverse Hessian ${\bf H}^{-1}$ of (\ref{eq: stage1_eqs}) evaluated at the solution $\Theta_{{\rm MAP}1}$~\cite{Petra:2017}.
\subsection*{Stage 2}
Stage 1 is effectively a Bayesian update for the generator parameters. In stage 2, the prior variances associated with the generator parameters in ${\Gamma}_g$ are set equal to the diagonal values of the inverse Hessian ${\bf H}^{-1}$ from stage 1, and the mean values of the generator parameters in $\overline{\Theta}_g$ are set equal to $\Theta_{{\rm MAP}1}$. Additionally, the full set of data in $\tilde{\bf V}$ and $\tilde{\bf I}$ is included in building the likelihood function, and the current injection variables are introduced into the framework. Again, current injections outside the range of $\Omega_d\pm\Omega_r$ may be neglected if it is known a priori that the FO is not occurring in these frequency bands. Otherwise, the optimization problem and solution in stage 2 are fully characterized by the formulation introduced in subsections \ref{Apply_MAP} and \ref{NH for MAP_Solution}. The value of $\lambda$ from~(\ref{eq: min_final}) should be set sufficiently high such that the optimizer finds a sparse set of current injection parameters. Although a cross-validation approach should typically be employed to choose this regularization parameter, our ultimate desire is to locate the sources of FOs rather than build the most accurate predictive model possible, meaning increased regularization may be permissible.

The full set of steps necessary to implement this FO source location procedure are outlined in Algorithm \ref{alg: FO}; several of these steps reference equations from~\cite{Chevalier:2018}. This algorithm concludes by comparing the size of the current injection solutions in ${\Theta}_{\mathcal I}$ to an operator specified threshold parameter $\iota$.

\begin{algorithm}\label{alg: FO}
\caption{Generator Source Detection Method}
\textbf{START}\\
\Repeat{completed for each of the \rm{g} \textit{generators}}{
\begin{enumerate}[label=\textbf{\arabic*},start=1]
    \item Analytically construct FRF $\mathcal{Y}$ of ~\cite[eq. (46)]{Chevalier:2018}\\
          via DAE sets~\cite[eq. (44)]{Chevalier:2018} and~\cite[eq. (45)]{Chevalier:2018}
    \item Take DFT of generator terminal PMU data ${\rm V}(t)$,\\
          $\theta(t)$, ${\rm I}(t)$, and $\phi(t)$ to yield ${\tilde {\bf I}}(\Omega)$ and ${\tilde {\bf V}}(\Omega)$
    \item Define parameter prior means and variances
    \item Use (\ref{eq: PMU_fft_noise}) and (\ref{eq: Nr})-(\ref{eq: Qi}) to build the likelihood covariance matrix of (\ref{Gamma2_{L}})
\end{enumerate}}

\vspace*{0.1cm}   
\textbf{Bayesian Stage 1}\\
\vspace*{0.1cm}

\begin{enumerate}[label=\textbf{\arabic*},start=7]
    \item Identify the range of frequencies $\Omega_d\pm\Omega_r$ where\\
    the forced oscillation has significant effect and\\ remove corresponding data from ${\tilde {\bf I}}(\Omega)$ and ${\tilde {\bf V}}(\Omega)$
    \item Iterate to local minimum ($\Theta_{{\rm MAP}1}$) of (\ref{eq: stage1_eqs}) 
        \begin{itemize}
        \item{Continuously update the covariance matrix at\\ each iteration, as described in subsection \ref{NH for MAP_Solution}}
      \end{itemize}
\end{enumerate}

\vspace*{0.1cm}   
\textbf{Bayesian Stage 2}\\
\vspace*{0.1cm}

\begin{enumerate}[label=\textbf{\arabic*},start=12]
	\item Update generator parameter prior means with\\
    $\Theta_{{\rm MAP}1}$ and prior variances with ${\bf H}^{-1}$ of (\ref{eq: stage1_eqs})
	\item Iterate to local minimum ($\Theta_{{\rm MAP}2}$) of (\ref{eq: min_sum}) while\\
    neglecting injection variables outside $\Omega_d \pm\Omega_r$
  	\item Via (\ref{eq: Theta_full}) and (\ref{eq: I_concat}), parse the stage 2 solution \\
        for generator $i$ injection vector $\boldsymbol{\mathcal{I}}_{i}=[\boldsymbol{\mathcal{I}}_{{\rm I}_{i}}^{\top}\;\boldsymbol{\mathcal{I}}_{\phi_{i}}^{\top}]^{\top}$
\end{enumerate}

\uIf{$\left\Vert\boldsymbol{\mathcal{I}}_i\right\Vert_\infty>\iota$\vspace*{0.1cm}}
{Source found at generator $i$}
\Else{\vspace*{0.1cm}No source found at generator $i$}
\end{algorithm}


\section{Test Results}\label{Test Results}
In this section, we test our FO source location method on two test cases. First, we consider a 4-bus power system (Fig. \ref{fig: Three_3rd_Order_+_Inf}), and we apply a sinusoidal oscillation to the mechanical torque supplied to one of the generators. Second, we apply two FOs to generators in the WECC 179-bus power system. In each test, white measurement noise is added to all PMU data (magnitude and phase) to achieve an SNR\footnote{In setting the SNR, the signal power of all angular data is found after angles are subtracted from the angle associated with the system's so-called center of inertia angle $\theta_{\rm coi}$ where $\theta_{\rm coi}=(\sum H_i\delta_i)/(\sum H_i)$~\cite{Milano:2013}.} of 45 dB in accordance with~\cite{Brown:2016}. For model explanation brevity, all simulation code has been publicly posted online\footnote{https://github.com/SamChevalier/FOs} for open source access.

\subsection{Three Generators Tied to an Infinite Bus}\label{3 Bus Test Results}
\begin{figure}
\begin{centering}
\includegraphics[scale=0.9]{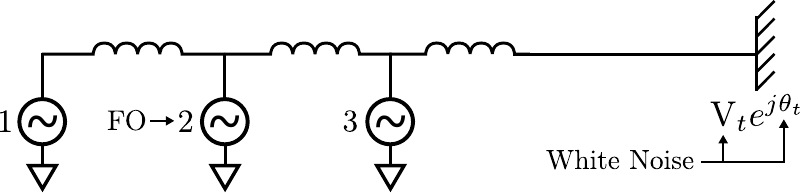}
\caption{\label{fig: Three_3rd_Order_+_Inf} Three $3^{\rm rd}$ order generators (with AVRs) are radially tied to an infinite bus with white noise. An FO is applied to generator 2's torque.}
\end{centering}
\end{figure}

In this test case, three $3^{\rm rd}$ order generators, each outfitted with first order automatic voltage regulators (AVRs), were radially tied to an infinite bus, as given by Fig. \ref{fig: Three_3rd_Order_+_Inf}. At the infinite bus, white noise was applied to simulate stochastic system fluctuations. The mechanical torque applied to generator 2 was forcibly oscillated via $\tau_m(t) = \tau_0(1 + 0.05\sin(2\pi0.5t))$.

\begin{figure}
\begin{centering}
\includegraphics[scale=0.45]{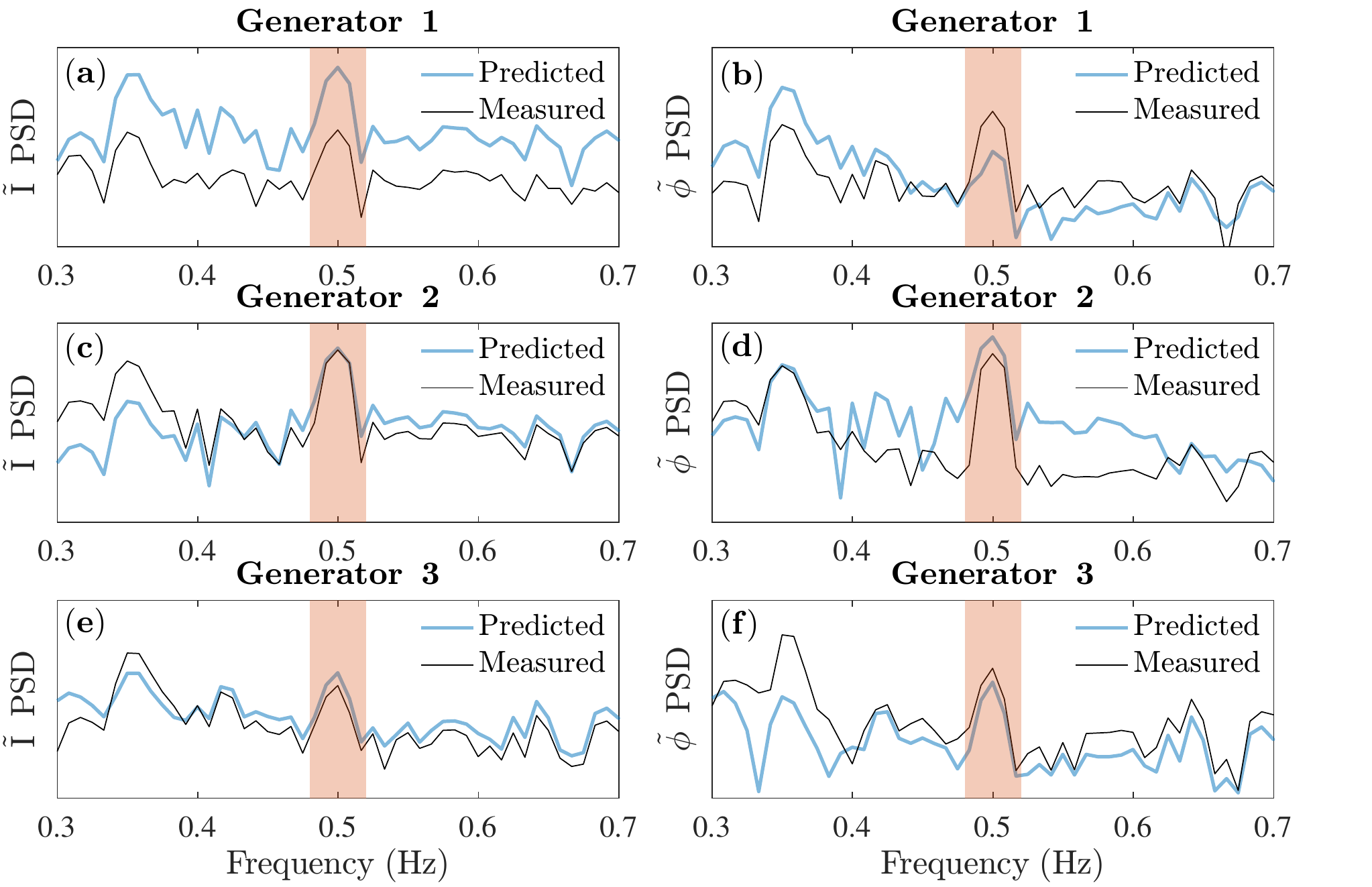}
\caption{\label{fig: 3_Gen_Initial} Shown are the measured (${\tilde {\bf I}}$) and predicted (${\mathcal Y}{\tilde {\bf V}}$) current magnitude (${\tilde {\rm I}}$) and current phase (${\tilde {\phi}}$) power spectral density (PSD) across a range of frequencies around the forcing frequency (0.5 Hz) before optimization has occurred. Panels $({\bf a})$ and $({\bf b})$ correspond to generator 1, panels $({\bf c})$ and $({\bf d})$ correspond to generator 2, and panels $({\bf e})$ and $({\bf f})$ correspond to generator 3.}
\end{centering}
\end{figure}

After 120 seconds of simulation, we built the generator admittance matrices using AVR and generator parameter ($D$, $H$, $X_d$, $X_d'$, $X_q$, $T_{d0}'$, $K_A$, and $T_A$) values which were numerically perturbed by a percentage value randomly chosen from $\mathcal{U}(-75,75)$. These perturbed parameter values represent the prior means placed into $\overline{\Theta}_g$ for MAP stage 1. Fig. \ref{fig: 3_Gen_Initial} compares the predicted current ${\mathcal Y}{\tilde {\bf V}}$ with the measured current ${\tilde {\bf I}}$ across a small range of frequencies before MAP is solved. At all generators, there is significant spectral deviation between the measurements and predictions. From this data alone, it is not clear which is the source generator of the FO.

\begin{figure}
\begin{centering}
\includegraphics[scale=0.448]{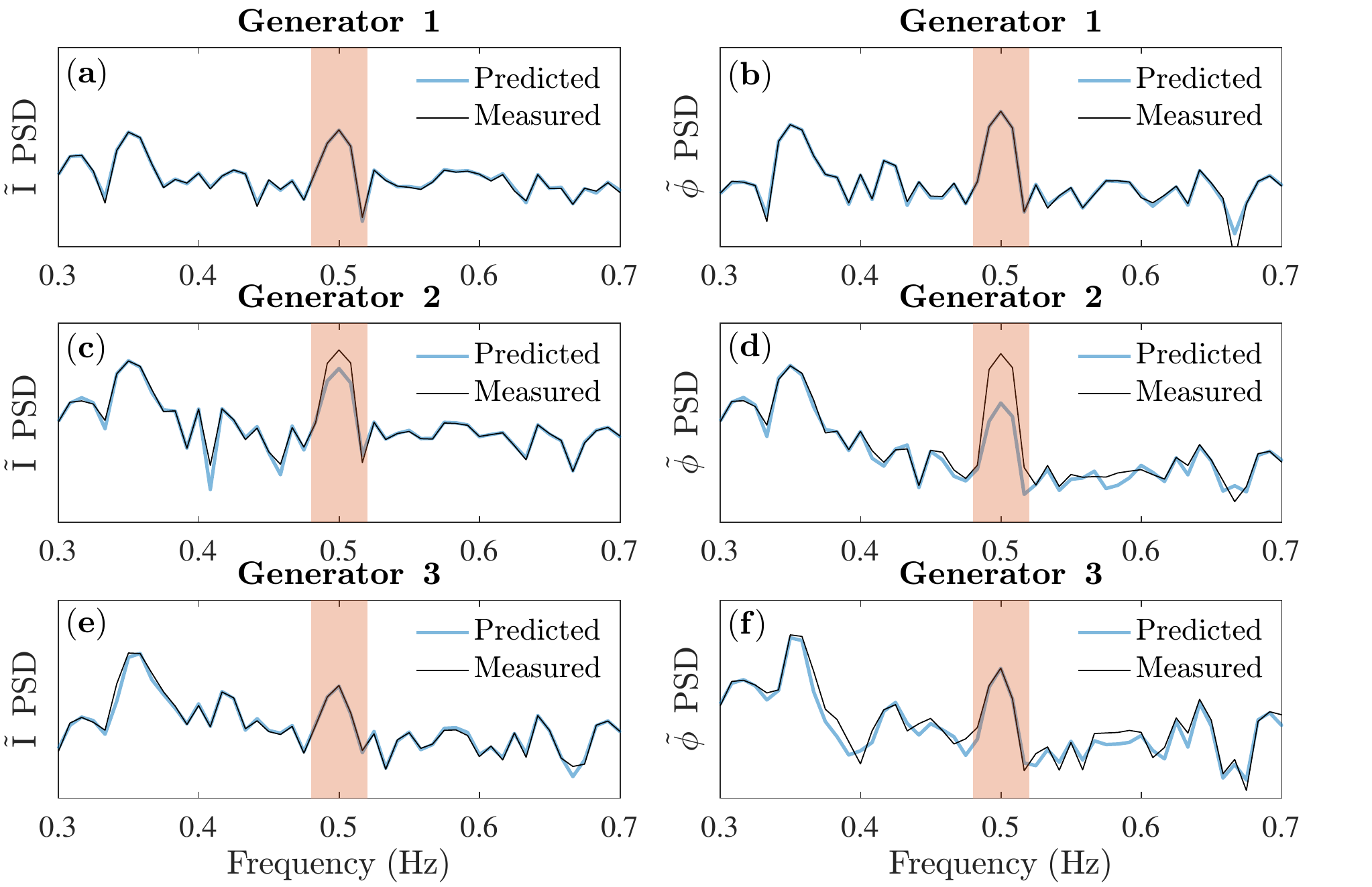}
\caption{\label{fig: 3_Gen_Final} Shown are the measured (${\tilde {\bf I}}$) and predicted (${\mathcal Y}{\tilde {\bf V}}$) current magnitude (${\tilde {\rm I}}$) and current phase (${\tilde {\phi}}$) power spectral density (PSD) around the forcing frequency (0.5 Hz) after stage 1 of the optimizer has been run.}
\end{centering}
\end{figure}

Next, stage 1 of the Bayesian update was run, where data in the range of the FO (the red shaded band in Fig. \ref{fig: 3_Gen_Initial}) was taken out of the problem altogether and current injections were not considered. After converging to $\Theta_{{\rm MAP}1}$, the new set of generator parameters was used to compute the predicted spectrums in Fig. \ref{fig: 3_Gen_Final}. Strong agreement between the measured and predicted spectrums is evident outside of the red band of the forcing frequency (which were not included in the stage 1 optimization).

\begin{figure}
\begin{centering}
\includegraphics[scale=0.452]{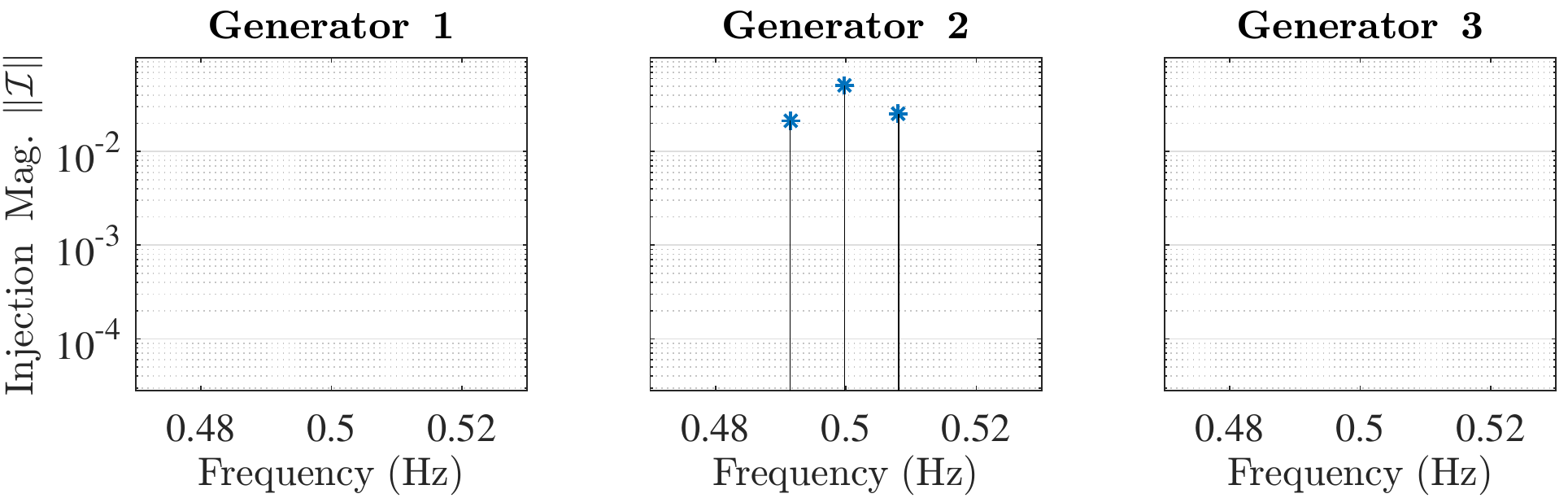}
\caption{\label{fig: I_injs} Shown are the stage 2 current injection results, as quantified by (\ref{eq: I_mag}).}
\end{centering}
\end{figure}

In running stage 2 of the optimization, the current injection variables were reintroduced and the full set of frequencies were optimized over. The results are summarized in Fig. \ref{fig: I_injs} which shows the norm of the current injections found at each frequency at each generator. For clarity, we plot 
\begin{align}
\left \Vert {\mathcal I} \right\Vert &= \sqrt{\mathcal{I}_{{\rm I}_{r}}^{2}+\mathcal{I}_{{\rm I}_{i}}^{2}+\mathcal{I}_{\phi_{r}}^{2}+\mathcal{I}_{\phi_{i}}^{2}}.\label{eq: I_mag}
\end{align}
In viewing these results, the size of the current injections identified by MAP at generator 2's forcing frequency of $0.5$ Hz are sufficiently large enough, when compared to the other generators (which are many orders of magnitude smaller), to clearly indicate the presence of a forcing function at this generator.

\subsection{Two Forced Oscillations in the WECC 179-bus System}\label{WECC 2 FOs}
In conjunction with the IEEE Task Force on FOs, Maslennikov \textit{et al.} developed a set of standardized test cases to validate various FO source detection algorithms~\cite{Maslennikov:2016}. For further testing, we applied the methods presented in this paper on data collected from a modified version of test case ``F1" in~\cite{Maslennikov:2016}, in which the Automatic Voltage Regulator (AVR) reference signal at a generator (generator 1) in the WECC 179-bus system was forcibly oscillated at 0.86 Hz. In modifying the system, a second FO of frequency 0.7 Hz was added to the mechanical torque of the generator (generator 15) at bus 65. Additionally, PMU measurement noise was added as described previously and Ornstein-Uhlenbeck noise (with parameters taken from \cite{Ghanavati:2016}) was added to all constant power loads, as described in \cite{Chevalier:2018}. After simulating the system, the admittance matrices for all system generators were constructed by parameters which were perturbed as in the previous subsection. For the second order generator model, parameters $H$, $D$, $X_d'$ and ${\rm E}'$ were perturbed. For the third order generator model, the time constant, reactance, inertia, and AVR gain parameters were perturbed.

\begin{figure}
\begin{centering}
\includegraphics[scale=0.435]{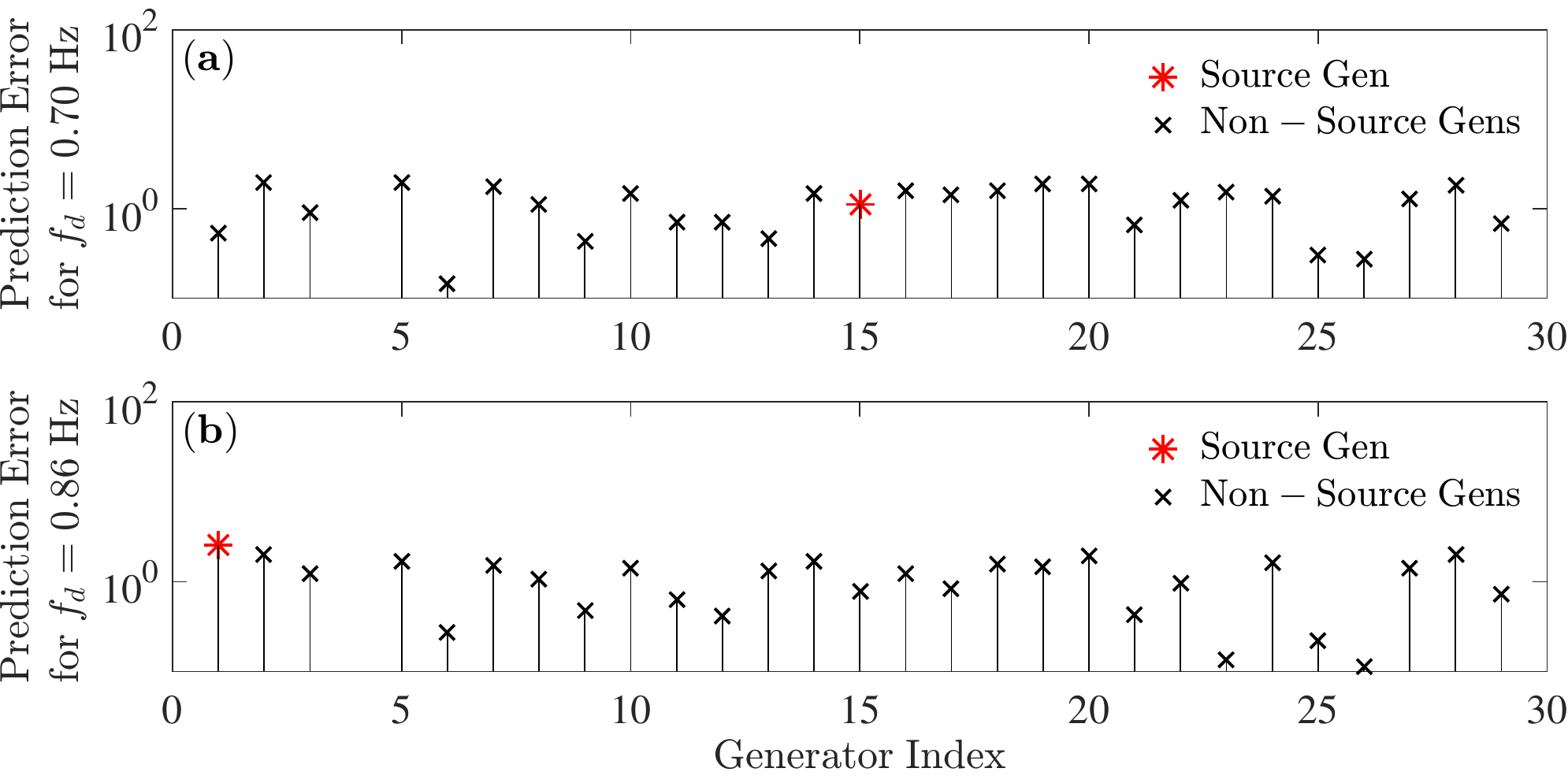}
\caption{\label{fig: I_error_179} The prediction error, as quantified by the percent difference in the measured and predicted currents in (\ref{eq: PD}), are shown for the system forcing frequencies of $0.70$ Hz in panel ($\bf a$) and $0.86$ Hz in panel ($\bf b$).}
\end{centering}
\end{figure}

Next, the measured and predicted spectrums were compared at both of the forcing frequencies of $f_d=0.70$ Hz and $f_d=0.86$ Hz. To visualize the initial prediction error, the \textit{percent difference} between measured and predicted currents are quantified via
\begin{align}
\label{eq: PD}
\text{Prediction Error \% Difference}\Rightarrow\frac{\left\Vert \tilde{{\bf I}}-\mathcal{Y}\tilde{{\bf V}}\right\Vert }{\frac{1}{2}\left\Vert \tilde{{\bf I}}\right\Vert +\frac{1}{2}\left\Vert \mathcal{Y}\tilde{{\bf V}}\right\Vert}
\end{align}
and plotted in Fig. \ref{fig: I_error_179}. The true source generator is identified in each panel, but because prediction error is sufficiently large due to parameter inaccuracies, it is not readily identifiable.

\begin{figure}
\begin{centering}
\includegraphics[scale=0.435]{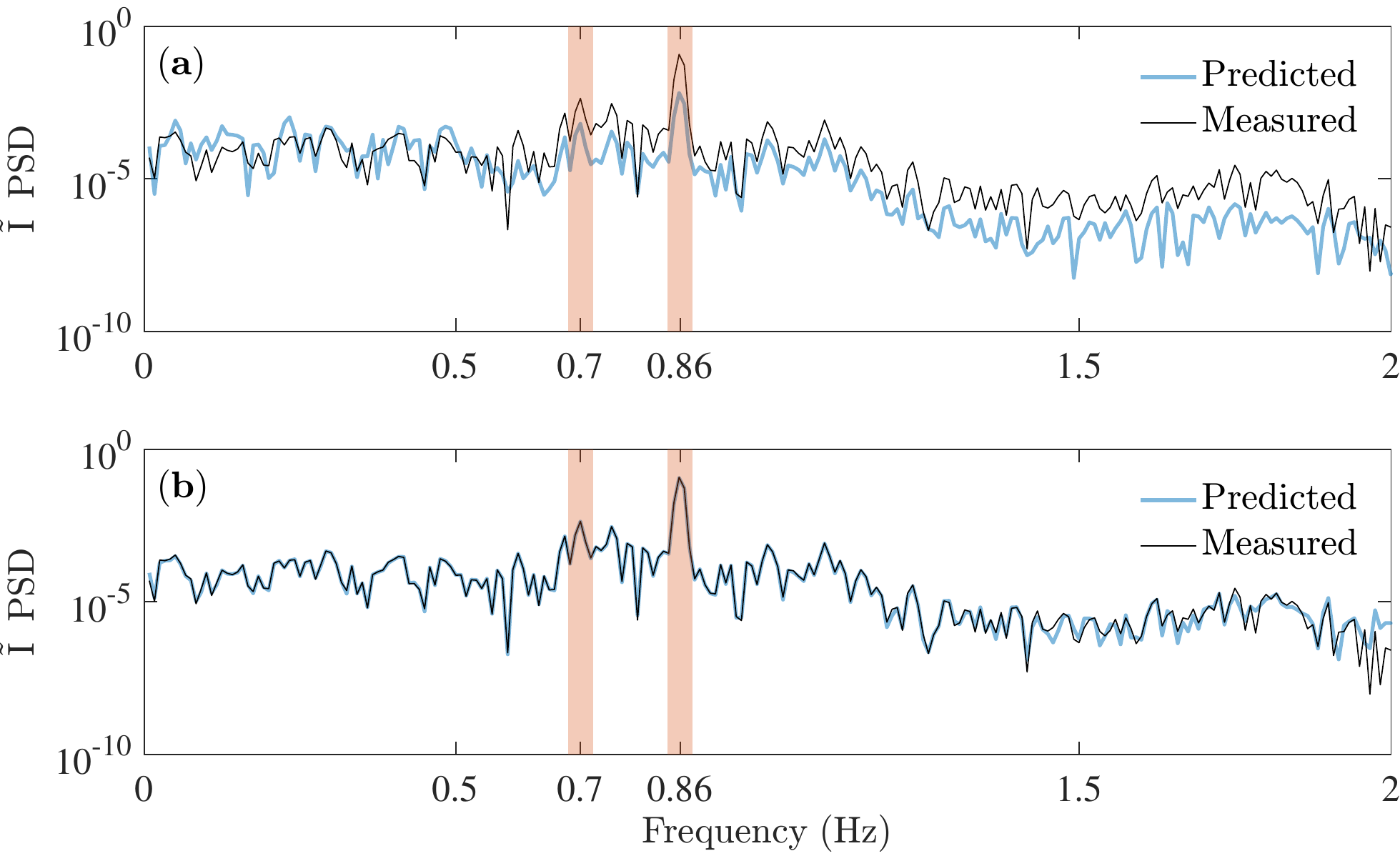}
\caption{\label{fig: Gen_2_Init_Final} The predicted current magnitude spectrum is plotted in panel ($\bf a$) for generator 3 (located at bus 9) before stage 1 of the optimization procedure. The predicted current magnitude spectrum is replotted in panel ($\bf b$) after stage 1 in completed. Generator 3 is not an FO source.}
\end{centering}
\end{figure}

\begin{figure}
\begin{centering}
\includegraphics[scale=0.435]{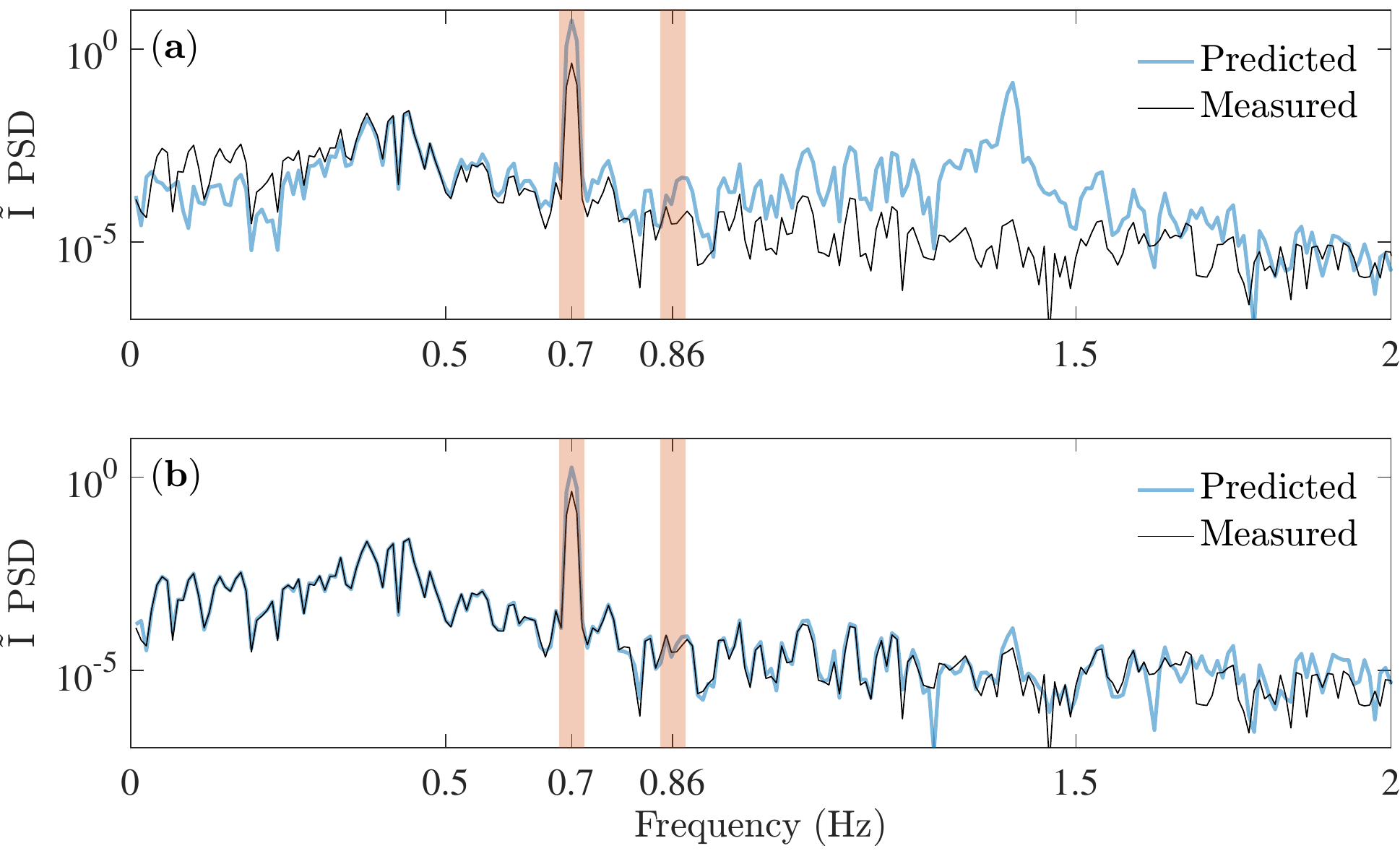}
\caption{\label{fig: Gen_14_Init_Final} The predicted current magnitude spectrum is plotted in panel ($\bf a$) for generator 15 (located at bus 65) before stage 1 of the optimization procedure. The predicted current magnitude spectrum is replotted in panel ($\bf b$) after stage 1 in completed. Generator 65 is an FO source at 0.70 Hz.}
\end{centering}
\end{figure}

Stage 1 of the algorithm was then run. The results are given for two representative generators: a non-source generator at bus 9 (Fig. \ref{fig: Gen_2_Init_Final}) and a source generator at bus 65 (Fig. \ref{fig: Gen_14_Init_Final}). Panel ($\bf a$) in Fig. \ref{fig: Gen_2_Init_Final} seems to indicate that generator 3 might be the source of the 0.86 Hz FO due to the large measurement/prediction deviations at this frequency, but the optimizer is able to reconcile the spectrums in panel ($\bf b$). Panel ($\bf b$) in Fig. \ref{fig: Gen_14_Init_Final}, however, shows a significant gap between the measured and predicted spectrums at 0.70 Hz caused by the FO. Due to the physically meaningful way in which the covariance matrix is constructed, the amplification of the measurement noise at the points of FRF resonance does not prevent the optimizer from converging to the true set of generator parameters, but the effect can become troublesome if the SNR of the PMU data drops too low.


\begin{figure}
\begin{centering}
\includegraphics[scale=0.435]{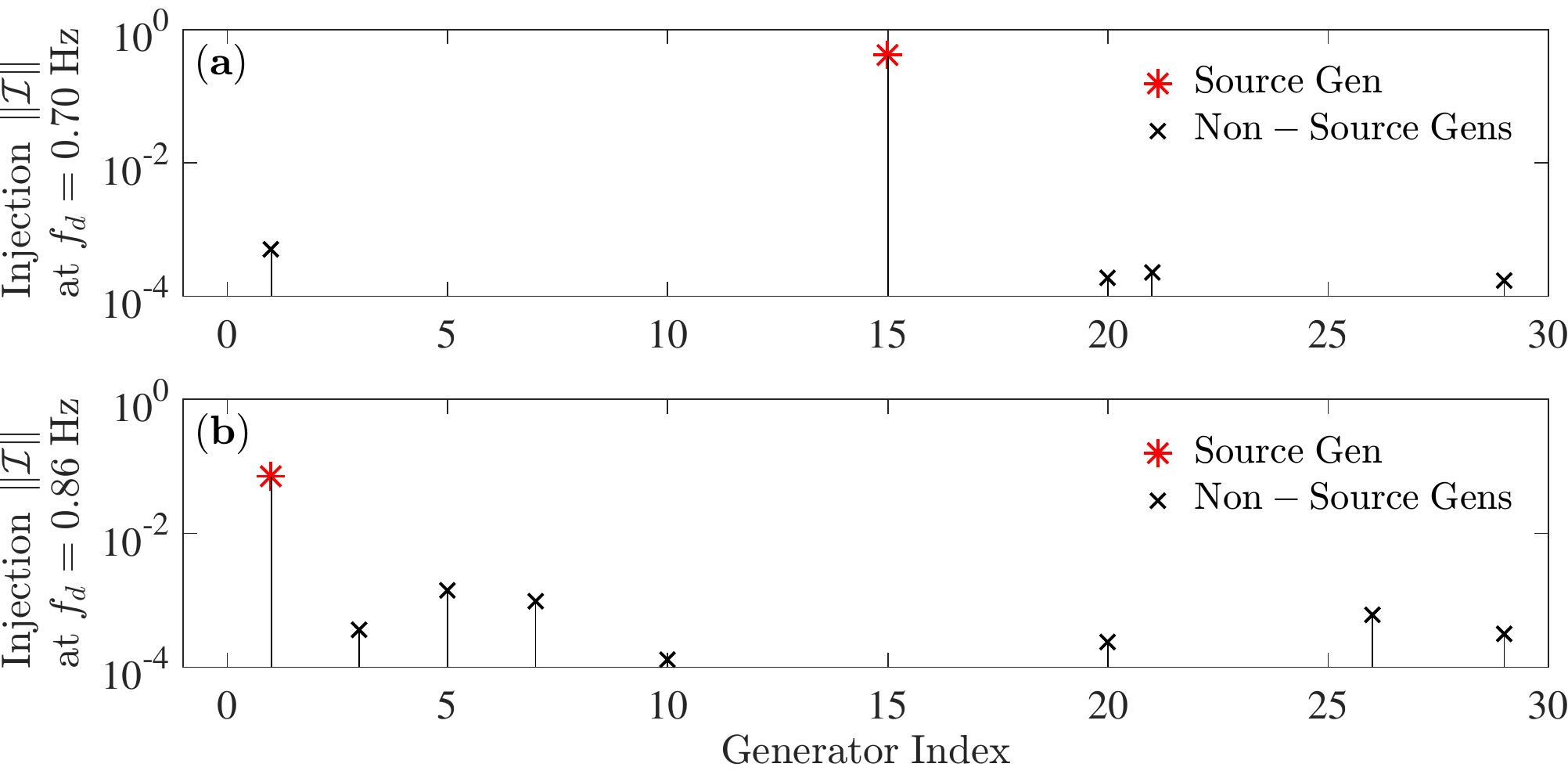}
\caption{\label{fig: Injections_179} Current injection magnitudes of (\ref{eq: I_mag}) found by the optimizer in stage 2 at both forcing frequencies.}
\end{centering}
\end{figure}

Finally, stage 2 of the optimization was run. Fig. \ref{fig: Injections_179} shows the magnitude of the current injections, as quantified by (\ref{eq: I_mag}), found by the optimizer. Although no threshold has been established, it is clear that generator 15 is the source of the $0.70$ Hz oscillation and that generator 1 is the source of the $0.86$ Hz oscillation. In general, as the PMU measurement SNR is driven higher, the injections found by the $\ell_1$ norm minimization in (\ref{eq: Theta_MAP}) at non-source generators are driven to 0.

\section{Conclusion and Future Work}\label{Conclusion}
In this paper, we have developed a Bayesian framework for locating the sources of forced oscillations in power systems assuming noisy PMU signals and uncertain generator parameters. In both of the provided test cases, the optimizer was able to solve for generator parameters with a sufficiently high degree of accuracy, and the origins of the FOs were successfully located with a high degree of certainty. Additionally, the method showed good performance in the context of a system experiencing multiple concurrent FOs.

Although applied exclusively to generators in this paper, we plan to extend these methods to other dynamic elements of the power system, such as dynamic loads or FACTS devices, which may also represent FO sources. Additionally, although Algorithm \ref{alg: FO} has been designed for the purpose of locating FO sources, it also presents an interesting method for performing system identification (SID), with direct applications to dynamic model verification and load modeling.
\section*{Acknowledgment}
The authors gratefully acknowledge Luca Daniel and Allan Sadun for their helpful suggestions on constructing the inverse problem and formulating the optimization framework.

\bibliographystyle{IEEEtran}
\bibliography{FO_Bib}

\begin{IEEEbiography}[{\includegraphics[width=1in,height=1.294in]{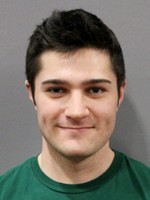}}]{Samuel C.~Chevalier} (S`13) received M.S. (2016) and B.S. (2015) degrees in Electrical Engineering from the University of Vermont, and he is currently pursuing the Ph.D. in Mechanical Engineering from the Massachusetts Institute of Technology (MIT). His research interests include power system stability and PMU applications.
\end{IEEEbiography}

\begin{IEEEbiography}[{\includegraphics[width=1in,height=1.294in]{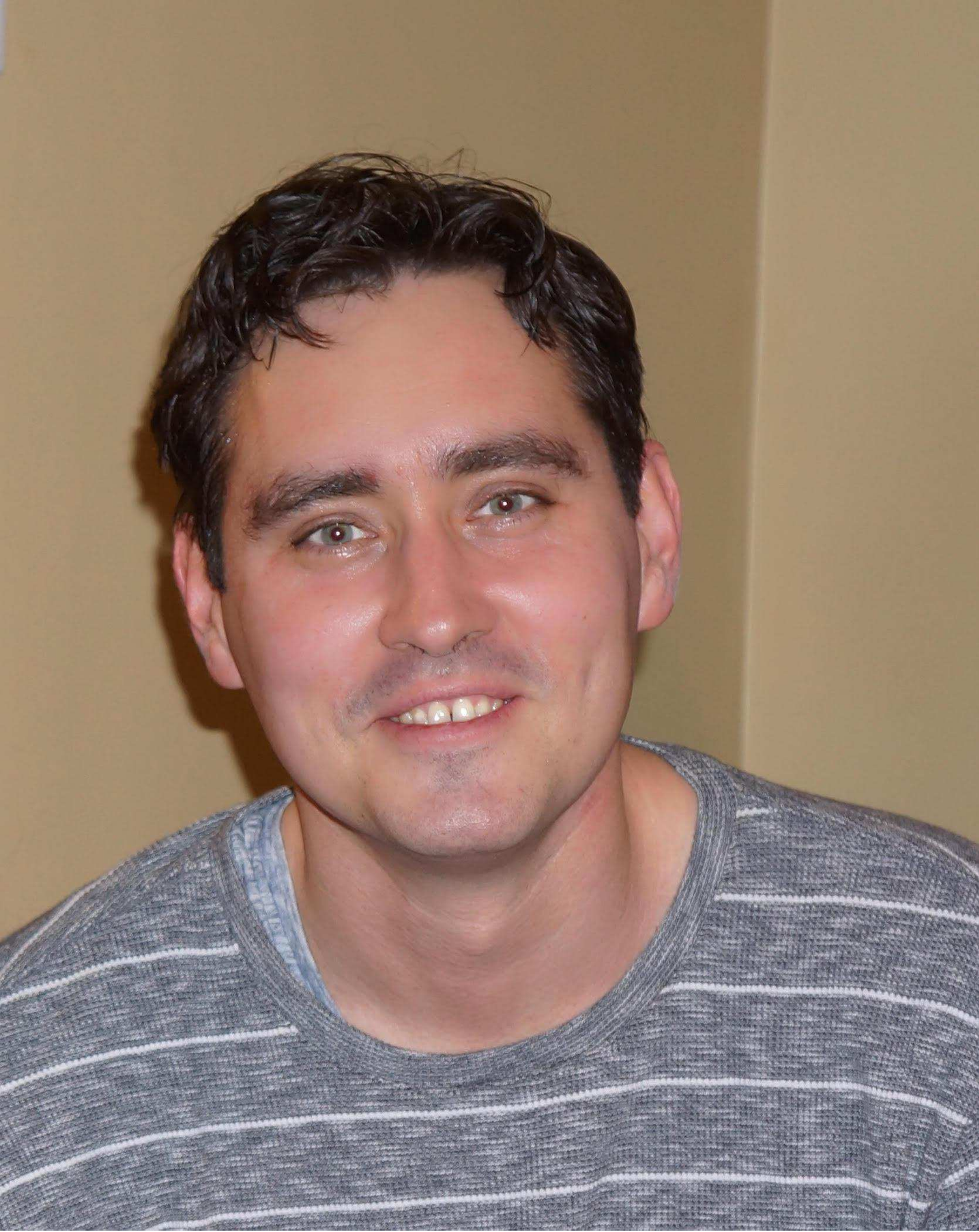}}] {Petr Vorobev} (M`15) received his PhD degree in theoretical physics from Landau Institute for Theoretical Physics, Moscow, in 2010.  From 2015 till 2018 he was a Postdoctoral Associate at the Mechanical Engineering Department of Massachusetts Institute of Technology (MIT), Cambridge. Since 2019 he is an assistant professor at Skolkovo Institute of Science and Technology, Moscow, Russia. His research interests include a broad range of topics related to power system dynamics and control. This covers low frequency oscillations in power systems, dynamics of power system components, multi-timescale approaches to power system modelling, development of plug-and-play control architectures for microgrids.
\end{IEEEbiography} 

\begin{IEEEbiography}[{\includegraphics[width=1in,height=1.294in]{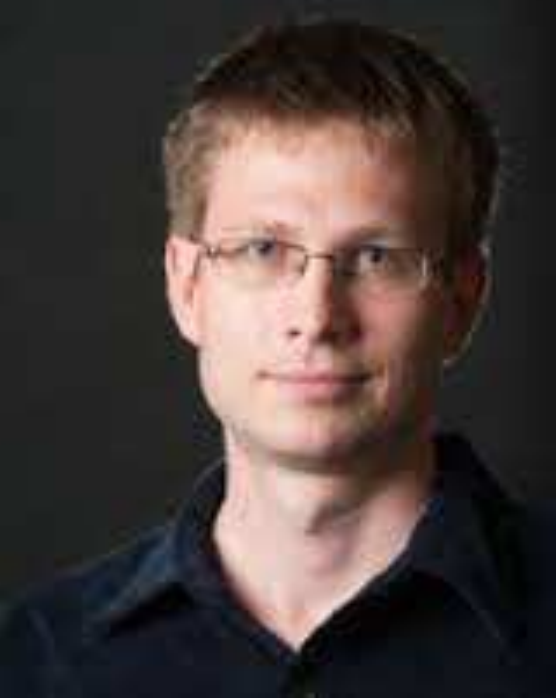}}]{Konstantin Turitsyn} (M`09) received the M.Sc. degree in physics from Moscow Institute of Physics and Technology and the Ph.D. degree in physics from Landau Institute for Theoretical Physics, Moscow, in 2007.  Currently, he is an Associate Professor at the Mechanical Engineering Department of Massachusetts Institute of Technology (MIT), Cambridge. Before joining MIT, he held the position of Oppenheimer fellow at Los Alamos National Laboratory, and Kadanoff-Rice Postdoctoral Scholar at University of Chicago. His research interests encompass a broad range of problems involving nonlinear and stochastic dynamics of complex systems. Specific interests in energy related fields include stability and security assessment, integration of distributed and renewable generation.
\end{IEEEbiography}

\end{document}